# How to face the loss in plasmonics and metamaterials

*Jacob B Khurgin*

Metal losses affect the performance of every plasmonic or metamaterial structure; dealing with them will determine the degree to which these structures will find practical applications.

**Loss happens**

The last decade and a half has seen a steady rise of interest in plasmonics and metamaterials (P&M), a duo of closely related fields. The inspiration driving P&M research is the enticing prospect of concentrating and manipulating the electromagnetic field on the sub-wavelength scale to achieve unprecedented enhancement of linear and nonlinear optical processes, and develope nano-scale optical interconnects, optical cloaking, and super-resolution imaging [1-3]. Inspired by a few visionaries, P&M research has been enabled by the determined work of an army of engineers and scientists in microelectronics research who have been making steady advances in nanofabrication. However, when it comes to the materials used in P&M, their palette has remained sparse and unchanged, relying mostly on the metals used in reflection optics, predominantly silver and gold. The reason for the choice of these two noble metals is obviously the low rate of scattering of free electrons in them, which leads to high conductivity in the RF range and high reflectivity in the optical range. And yet, after decade-long effort, which did produce attention-grabbing results, the P&M community is beginning to face the unpleasant fact that, even in the noblest of metals, losses in the optical range (and here we include near- and mid-infrared regions) are still too high to make most, if not all, practical P&M devices. At this time, the one and only wide-spread practical application of plasmonics is surface enhanced Raman scattering (SERS) [4] which predates the term "plasmonics" itself, whereas the best results in nano-antennas and metamaterials have been attained in the microwave range, where these structures have been known, albeit under a different nom-de-guerre "artificial dielectrics" [5], since the early 1950s.

If one is to summarize the current state of efforts to enhance performance of various optical devices using P&M, it can be done by noticing that while inefficient devices can indeed be improved, the performance of any relatively efficient device will only deteriorate, as loss in the metal will exceed whatever improvement the field concentration achieves. That is why for fluorescence sensors, where the absolute efficiency (light out vs. light in) is typically quite low, P&M does offer some promise. (Raman scattering can be thought of as fluorescence with efficiency approaching zero – hence the aforementioned spectacular success of SERS!) But when it comes to the devices where the efficiency must be high such as light sources and detectors, solar cells, interconnects, switches and modulators, the community is gradually beginning to realize that unless metal loss is reduced significantly the promise of P&M will remain unfulfilled [6,7].

A number of strategies for loss mitigation (or avoidance) are being pursued. These involve using advances in fabrication aimed at reduction of surface roughness and effects [8], using highly doped semiconductors in place of metals [9], using polar dielectrics in the Reststrahlen region (phononics) [10], and using media with optical gain. Furthermore, a number of groups have realized that if absorption in the metal cannot be avoided, it may as well be used for productive purposes, in applications ranging from photo-detection [11] to solar photocatalysis [12]. In this brief Commentary I will try to reflect on the prospects of each of these directions, in the process making a few observations of fundamental nature.

**Why does plasmonics always mean loss?**

First of all, it is important to emphasize that metal (or any material with negative permittivity $\varepsilon$ ) is essential for field concentration beyond the diffraction limit. Let us consider an optical resonator with characteristic dimension *a* larger than the wavelength $\lambda$ in the dielectric (Fig.1a). When the electro-magnetic mode oscillates inside it, every half period the energy is transferred from the electric field energy $u_E \sim \frac{1}{2}\varepsilon E^2$ to the magnetic field energy $u_H \sim \frac{1}{2}\mu H^2$ and back. This is similar to the mechanical analog of an oscillating mass on a spring, where potential energy is transferred into kinetic energy and then back. Both *E* and *H* have harmonic dependence on the spatial coordinate *z*, as $\sin(2\pi z/\lambda)$. From the Maxwell equation $\nabla \times \boldsymbol{H} = \varepsilon \partial \boldsymbol{E}/\partial t$, it follows that $H = \sqrt{\varepsilon/\mu}E$ , and hence $u_H = u_E$, means that energy is conserved while transferred from one form to another. In this configuration self-sustaining oscillations are possible. Unsurprisingly, the second Maxwell equation $\nabla \times \boldsymbol{E} = -\mu \partial \boldsymbol{H}/\partial t$ is also automatically satisfied when $H = \sqrt{\varepsilon/\mu}E$. Now, if the characteristic size is less than $\lambda/2n$, where *n* is the refractive index, (Fig.1b) the field spatial dependence becomes $\sin(\pi z/a)$, and from the first Maxwell equation it follows that for truly sub-$\lambda$ mode, the magnetic energy is much smaller than the electric energy, which makes self-sustaining oscillations impossible, as expected from diffraction limit. If, however, one introduces free carriers into the mode (Fig.1c), the energy can be also stored in the form of kinetic energy $u_K \sim \frac{1}{2}\varepsilon_0 \left( \omega_p^2/\omega^2 \right) E^2$ (here $\omega_p$ is the plasma frequency that for noble metals lies in the UV). Thus, depending on the exact geometry, the energy balance $u_M + u_K = u_E$ can be restored at some frequencies $\omega_{spp}$, typically on the scale of a fraction of $\omega_p$), precisely the frequencies of sub-$\lambda$ surface plasmon polaritons (SPP). SPPs can be either localized or propagating along the metal/dielectric interface with effective wavelength that is much shorter than $\lambda/2n$. Note that since the magnetic field is small, the second Maxwell equation $\nabla \times \boldsymbol{E} = -\mu \partial \boldsymbol{H}/\partial t$ dictates that the electric field must be nearly irrotational, resembling an electro-static field; hence the term "quasi-static" that is often use to describe sub-$\lambda$ regime.

Note that in principle one could have also approached the sub-wavelength confinement from the opposite angle, by considering the situation where $E$ is small, $E \ll \sqrt{\mu/\varepsilon}H$, while $H$ is quasi-static and irrotational, and $u_E \ll u_M$. Then to restore energy balance one would have to find a way to store additional potential energy, which cannot be done without increasing the dielectric constant and thus going back to the diffraction limit. Whereas free carriers naturally have large kinetic but no potential energy, finding an entity with the opposite characteristics at optical frequencies is not easy. In other words, an electric dipole is perfectly capable of oscillating without substantial magnetic field, but a magnetic dipole always requires electric field somewhere in its vicinity in order to support self-sustaining oscillations.

Based on this reasoning, one can also choose to combine two separate regions – one storing electrical energy and the other storing magnetic energy – into a single cell of metamaterial, essentially creating LC resonance circuits on a microscopic level (a split ring resonator [13] is one example). But as dimensions of sub-wavelength inductors become comparable with the plasma wavelength $\lambda_p = 2\pi c/\omega_p \approx 200 nm$, corresponding to an operational wavelength of a few micrometers or less, most of the energy will be stored in the form of kinetic energy of electrons rather than as magnetic energy [6], which brings us back to the situation similar to Fig1c. Or, in other words, in structures that are comparable to the plasma wavelength in at least one dimension the kinetic inductance exceeds the magnetic inductance. . Only at wavelengths longer than a few micrometers, a smaller proportion of energy gets stored in the kinetic motion of electrons leading to lower losses. That is why nano antennas and metamaterials in the mid-IR and longer wavelength range operate with much smaller loss than in optical and near-IR range.

So, the diffraction limit can be beaten with the help from free carriers, but this success comes at a price – almost half of the time the energy is stored in the form of kinetic energy, and therefore it is lost at the rate that is commensurate with the rate of scattering of electrons in metal, $\gamma$ which is on the scale of 10 fs (more about that value later). Thus, once the optical mode becomes sub-$\lambda$ in all 3 dimensions, independent of its shape, the SPP energy will be lost on the 10 fs scale [6].

It is instructive to relate losses and the Purcell effect observed in P&M structures and used extensively to enhance fluorescence and Raman scattering. The Purcell effect follows from Fermi's Golden rule and states that the rate of spontaneous decay of excited atoms and molecules is proportional to the density of states into which emission takes place. The density of photons in free space is low because of large speed of light, but in the P&M structures, it is much higher and the rate of spontaneous emission can be increased manifold. One must recognize though, that the "new" states into which the emission takes place cannot magically appear out of nowhere: essentially they are the states (degrees of freedom) of the motion of free electrons that couple into the original photon states. Their density is very high (since the electron velocity is much less than *c*) but they also dissipate energy on the femtosecond scale. Hence Purcell enhancement in P&M structures is obtained by coupling energy into lossy electronic states, and once there, the energy has a much higher chance to be dissipated rather than radiated into space. That is why, as mentioned above, plasmonic enhancement works well only for originally weak and inefficient optical processes, such as Raman.

**Nuts and bolts of absorption in metals**

Let us now see what makes absorption in the metal so strong in the first place. Although most of the aspects of it was investigated thoroughly half-a-century ago [14], the P&M community has not paid enough attention to them. Consider the simplified band structure of a metal in Fig. 2 showing two states with wave-vectors (momenta) $k_1$ and $k_2$ and energies $E_1$ and $E_2 = E_1 + \hbar\omega$, respectively below and above the Fermi level. The magnitude difference between these two momenta is $\Delta k_{12} > \Delta k_0 \sim \omega/v_F$ (where $v_F \sim 10^8 cm/s$ is Fermi velocity), typically more than $3 nm^{-1}$, way too large to be supplied by a photon, or in our case a SPP. Therefore, the absorption process requires a phonon or an imperfection (Fig. 2a) and it results in two "hot" carriers – electron and hole -- whose energy can be harvested if and when they reach the surface of the metal. The probability of absorption assisted by phonons and defects is proportional to $\gamma_{ph}(\omega)\omega_p^2/\omega^2$ and weakly-dependent on frequency (typically, $\gamma_{ph}(\omega) \sim 10^{14} s^{-1}$). The reason for this large value, as follows from Fermi's golden rule, is a huge density of states above Fermi level available for the transition.

The other momentum-conserving process that can lead to absorption is electron-electron (e-e) scattering in which two electrons make the transition from the states $k_1$ and $k_2$ below the Fermi level to the states $k_3$ and $k_4$ (Fig.2b). The momentum conservation in this so-called Umklapp process $k_3 + k_4 = k_1 + k_2 + G$, where $G$ is the reciprocal lattice vector. The energy conservation relation $E_2' + E_1' = E_1 + E_2 + \hbar\omega$ indicates that rather than two "hot" carriers sharing energy $\hbar\omega$, one ends up with four "lukewarm" carriers having kinetic energies on the order of $\hbar\omega/4$ each – and this is the fact with the gravest of consequences for such applications as photodetection and energy harvesting in which hot carriers must eventually surpass some energy barrier. The e-e scattering is strongly frequency dependent since high photon energy means more electrons and holes involved in the process, and one can roughly say that $\gamma_{ee}(\omega) \sim \gamma_0(\hbar\omega/E_F)^2$ [14] where $\gamma_0 \sim 10^{15} s^{-1}$ [15] becomes comparable to $\gamma_{ph}$ in the visible range. Note that neither $\gamma_{ee}(\omega)$ nor $\gamma_{ph}(\omega)$ vanishes at low temperatures, because in the first case, no phonons participate in the electron-electron process, while in the second case, phonons can always get spontaneously emitted. Unlike the case of RF, any attempt to reduce the loss in optical range by cooling is therefore not advisable.

Another, often overlooked, mechanism that may lead to the absorption is associated with the finite size of SPP modes. If the effective size of the SPP mode inside the metal $a_m \ll \lambda/2$, it will include spatial frequencies larger than $\Delta k_0$, and therefore will introduce absorption by the "diagonal" process shown in Fig.2c. This process, known as Landau damping, leads to an additional "confinement" contribution to the absorption $\gamma_c \sim v_F/a$ which becomes comparable to the other "bulk" contributions $\gamma_{ph}$ and $\gamma_{ee}$ when $a \sim 10 nm$. This result has two important consequences for sub-$\lambda$ SPP modes. First, the "bulk" contribution plays a progressively less and less important role; even if an "ideal" metal with no bulk losses existed, confinement-induced absorption would still make losses enormous. Because there are plenty of available states, electrons will still make the transition; the only way to avoid it is to have a structure with a bandgap. Second, the confinement-dependent loss actually prevents a tight nanofocusing of light and limits the maximum attainable Purcell factor – an effect phenomenologically described as "diffusion" in the hydrodynamic model used in [16].

Finally, in all metals, there is band-to-band absorption (Fig. 2d), which becomes significant around 600 nm for Au, 400 nm for Ag, and in the UV for Ga. Note that when band-to-band absorption takes place, no hot carriers are formed, because the hot electron is created in the vicinity of the Fermi level, and while the hole in the d-band does have high energy, its velocity is very low because the effective mass in the d-band is very high.

Out of four mechanisms causing absorption in metals only two (a and c) actually generate hot carriers. Ironically, it is at higher photon energies, where one would want to have them that hot carriers are generated with relative low probability, as mechanisms b and d become the dominant ones.

Once the hot carriers have been generated, they lose their energy in two steps. First, e-e scattering in which a "hot" electron with energy $E_0$ above the Fermi level "collides" with one with energy below the Fermi level and as a result three carriers (one hole and two electrons) are generated with average energy of $E_0/3$. This process quickly leads to thermalization of hot carriers where they all have thermal energy $k_B T_e$ only slightly above the Fermi level. Just like e-e assisted absorption, the thermalization rate has quadratic energy dependence $\gamma_{th}(E) \sim \left[(E-E_F)/E_F\right]^2$ [15], but since it does not involve the optical field, thermalization does not have to be an Umklapp process. Hence the rate $\gamma_{th}$ is typically somewhat faster than $\gamma_{ee}$, and is on the scale of a $(10\text{fs})^{-1}$ for $E > 1eV$, as measured in [17]. Assuming longer thermalization times (up to 1ps), as made by many in the P&M community, is therefore rather unsubstantiated. Following thermalization, the electron temperature $T_e$ gets equalized with the lattice temperature $T_L$ via emission of acoustic phonons in the 100fs timescales. For prospective applications in energy harvesting and detections, requiring a certain threshold energy on the 0.5-1eV scale, one can safely say that hot carriers disappear after a few tens of fs, i.e. the time comparable to the scattering times $\gamma^{-1}$ in the Drude theory.

**What can and cannot be done?**

Let us now briefly survey the methods proposed for the reduction of loss in P&M. One approach is to use highly doped semiconductors in place of metal [9]. The rate of scattering in highly doped semiconductors is on the scale of $10^{13}\text{s}^{-1}$, simply because the Fermi level is much lower than in metals, and therefore the density of states into which the photo-excited carriers can end up is an order of magnitude lower than in metals. But $\omega_p$ in semiconductors is also an order of magnitude lower, making these materials most useful at mid-IR and longer wavelengths, where, for the most part, losses are also not that bad. Indeed, in this wavelength range, highly doped semiconductors have not shown performance superior to metals. Similarly one could consider superconducting P&M devices, but they would have to operate at energies less than the superconductive gap, i.e. in the THz or far IR, where, as we have already mentioned, metals can perform adequately.

If we now return to the argument made earlier about the necessity of having kinetic motion to achieve energy balance in sub-λ modes, note that the charged particles undergoing this motion do not have to be free electrons, and therefore may include ions in polar crystals, such as SiC, BN and others [10]. The dielectric constant [Fig.3] of the polar crystal can be written as $\varepsilon(\omega) = \varepsilon_\infty [1 + (\omega_{LO}^2 - \omega_{TO}^2)/(\omega_{TO}^2 - \omega^2 - i\omega\gamma)]$ where $\varepsilon_\infty$ is the dielectric constant at high frequencies, and $\omega_{TO}, \omega_{LO}$ are the frequencies of transverse and longitudinal optical phonons, laying in the mid-IR between 8 and 15μm. In the so-called Reststrahlen region $\omega_{TO} < \omega < \omega_{LO}$, the real part of ε is negative. Hence one can achieve sub-λ confinement of mid-IR radiation with relatively low loss, since the damping rate $\gamma \sim 10^{12}$ s$^{-1}$ – two orders of magnitude less than in metals. This fact makes "phononics", – the use of surface phonon polaritons (SPhP) – an attractive alternative to plasmonics. However, relying on phonons does have a significant drawback. As shown in Fig.3, in addition to kinetic energy $u_K(t)$, the oscillating ions also carry potential energy $u_P(t)$ which in the Reststrahlen region is only slightly less than $u_K(t)$. Therefore energy is transferred back and forth between $u_K$ and $u_P$ with only a small fraction of energy actually being that of the electric field. Formally this follows from the much stronger dispersion of $\varepsilon(\omega)$ in polar dielectrics than in metals, since the total energy is proportional to $\partial(\omega\varepsilon)/\partial\omega$.. In other words, in the vicinity of resonance, the energy is stored in the ion vibrations and not in the electric field. The net result is that one can attain a very high degree of concentration of energy in phononic materials, and narrow resonance line width, but when it comes to the electric field, the field concentration and Purcell factor are not much better than in metals. When it comes to propagating SPhPs, their lifetime is orders of magnitude longer than in SPPs, but the propagation length of SPhPs is not much longer than that of SPPs, since the group velocity of SPhP is very low due to high dispersion. To summarize this part, phononic materials are an attractive, relatively low loss sub-λ energy storage media, but they do not couple well with electromagnetic waves, while plasmonic materials do couple well with electromagnetic waves, but lose energy very fast. In my view, combining phononic structures with plasmonic nano-antennas may offer an attractive solution in the mid-infrared region of the spectrum.

We finally turn our attention to the often discussed idea of compensating metal losses in P&M structures with optical gain [1]. Although gain can be achieved in principle by optical pumping, practical devices usually require electrical pumping,. Even though the loss coefficient in SPPs approaches thousands of cm$^{-1}$, gain in semiconductor media can be just as high and transparency achieved when the carrier concentration is on the order of $N_{tr} \sim 10^{19}$cm$^{-3}$. This has led to extensive efforts to develop such loss-mitigated structures. Conceivably, what has been overlooked is that from the practical point of view what matters is the pump current density at which transparency is achieved, $i_{tr} \sim eN_{tr}d/\tau_{sp}$, where $d$ is the active layer thickness (~100 nm) and $\tau_{sp}$ is Purcell-shortened spontaneous recombination time. For tightly confined SPP, the Purcell factor easily surpasses 100; as a result, the transparency density exceeds 1MA/cm$^2$, [18], which is at least 1000 times higher than threshold current densities in conventional injection lasers. Such high pumping densities have been achieved by short pulse optical pumping, which always creates exciting news, but does not, in my opinion, lead to remotely practical devices.

**Parting thoughts**

In summary, I want to stress that the goal of this Commentary is not to offer a broad 'vision' of the future P&M developments as way too many competing (and compelling) visions have already propagated in the literature by the scientists who have mastered this art much better than I can ever hope. My modest objective was to present a few hard facts, substantiated by numbers, and then let the readers decide which of the visions can actually become a reality in the not-so-distant future. And yet, having painted such a -- let us say -- less than breathtakingly sunny landscape, I do feel compelled to enliven it with a few bright strokes where I can glimpse some glimmer of light in the P&M future. First of all, one would be well justified to focus activities in the mid-IR region of the spectrum, where the losses, be it metal, semiconductor or polar dielectric, become manageable, and where one can find numerous applications in thermal management and vibrational spectroscopy. When it comes to near-IR and visible spectral regions, with the materials available today, I simply cannot see any application where a decent absolute efficiency is required, be it in light emission, transmission, manipulation, detection, or harvesting that can benefit from incorporating P&M structures. On the other hand, sensing, where there is no need for high absolute efficiency, can be enhanced with P&M, although one needs to be very careful and not to forget about numerous competing methods of enhancing sensitivity, such as, for instance, micro-resonators [19], which, while perhaps not being as trendy as P&M, may offer equal if not better performance. And of course, one can always use high absorption in P&M to generate highly localized heat. The most exciting and paradigm-changing direction of P&M research, and I hope not just in my view, is search for novel materials with lower loss. This search, however, should not follow today's pattern of rapid-fire testing of all the well-known conductors, doped semiconductors, or popular materials-du-jour (graphene, MoS$_2$, and whatever comes next), in hope of a miracle, but instead should be a well thought-up and concerted effort by condensed matter theorists, chemists and growth specialists to synthesize man-made negative $\varepsilon$ materials with reduced loss [20]. With that, I believe, P&M science will live up to its promise.

*Jacob B Khurgin is at the Johns Hopkins University, Baltimore MD 21218, USA*

*e-mail: jakek@jhu.edu*

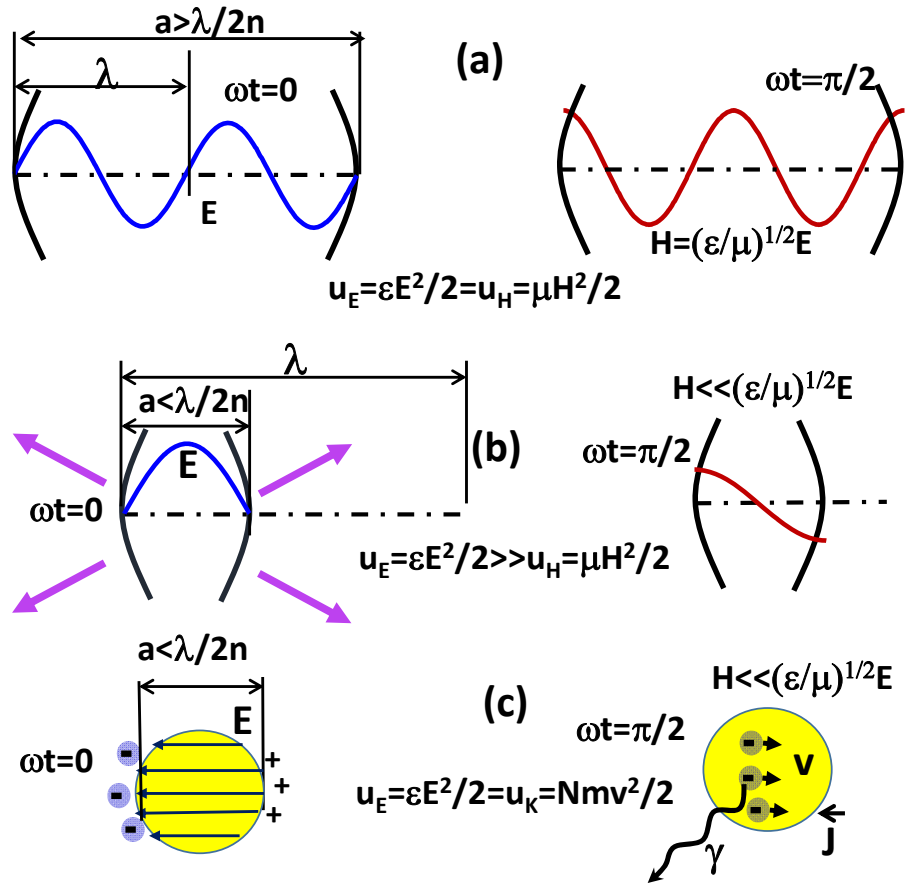

**Figure 1. Energy balance in photonic and plasmonic structures. (a)** In an optical cavity with dimensions larger than half wavelength $\lambda/2n$ energy alternates between $u_E$ (left), the energy of electric field $E$ and $u_M$ (right), energy of magnetic field $H$, analogous to energy alternating between potential and kinetic forms for the mass $m$ oscillating on the spring with velocity $v$. Here $\omega$ is angular frequency, $\varepsilon$ is electric permittivity, $\mu$ is magnetic permeability, $n$ is refractive index and H is the magnetic field. **(b)** In a sub-wavelength cavity with characteristic size $a$ magnetic energy is too small, hence the cavity radiates energy out-diffraction limit. **(c)** When free carriers are introduced and the current $J$ flows, the energy alternates between $u_E$ and kinetic energy of carriers $u_K$ where $N$ is the carrier density. The diffraction limit is beaten, but the motion of carriers is strongly damped with the damping rate $\gamma$, and the SPP mode is lossy.

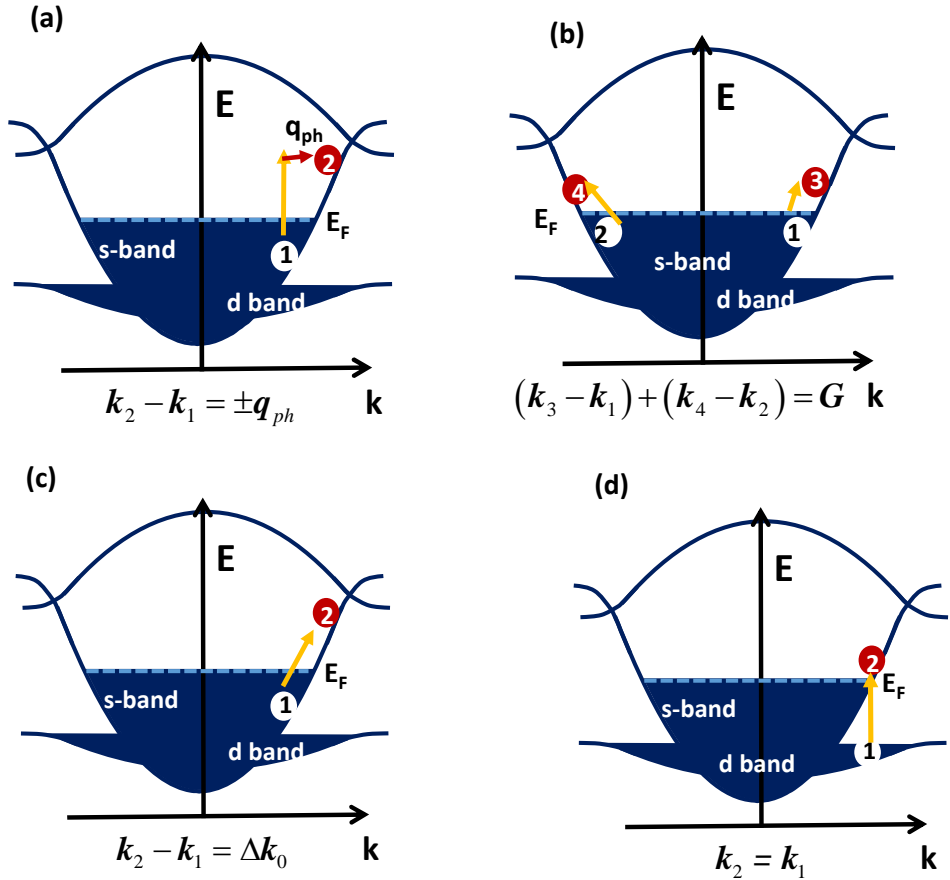

**Figure 2. Absorption of a quantum of electro-magnetic energy ℏω in a metal. (a)** absorption assisted by a phonon with wavevector $q_{ph}$ creates hot hole (1) with wave vector $k_1$ and hot electron (2) with wavevector $k_2$ and energy E above the Fermi level $E_F$ **(b)** e-e scattering assisted absorption, an Umklapp process involving reciprocal lattice vector $G$ creates four "lukewarm" carriers –two holes (1,2) and two electrons(2,3) **(c)** direct absorption (Landau damping) assisted by the SPP momentum $\Delta k_0$ creates hot hole (1) and electron (2) **(d)** interband absorption from d to s shell does not create hot carriers as hole (1) in d-band has low velocity and excited electron (2) is too close to Fermi level.

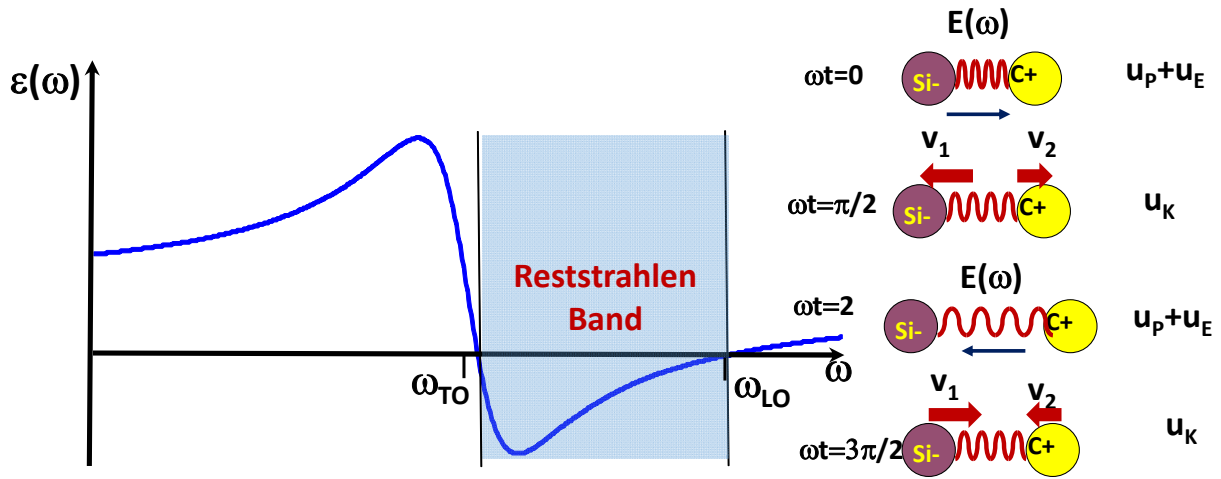

**Figure 3. Dispersion of the real part of the dielectric permittivity ε of polar crystal and the alternation of energy in it using SiC as an example.** $\omega_{TO}$ and $\omega_{LO}$ are the angular frequencies of transverse and longitudinal optical phonon modes, $v_1$ and $v_2$ are the velocities of ions in the presence of optical field $E(\omega)$, $u_P$ is a potential energy, $u_K$ is a kinetic energy and $u_E$ is the energy of electric field.